\documentstyle[prl,aps,floats,epsf,color,graphicx]{revtex}
\begin{document}
\baselineskip=12pt
\def\be{\begin{equation}}
\def\ee{\end{equation}}
\def\ba{\begin{eqnarray}}
\def\ea{\end{eqnarray}}
\def\la{\langle}
\def\ra{\rangle}
\def\a{\alpha}
\def\b{\beta}
\def\m{\mu}
\def\n{\nu}
\def\h{\hskip 1cm}
\def\hh{\hskip 2cm}
\def\lo{\longrightarrow}
\def\ro{\rightarrow}
\twocolumn[\hsize\textwidth\columnwidth\hsize\csname@twocolumnfalse\endcsname

\title { Exact ground states for a two parameter family of spin 1/2 ${\rm xyz}$ Heisenberg chains}

\author {M. Asoudeh$^{a}$, V. Karimipour$^{a}$, and A. Sadrolashrafi$^{a}$}

\vskip 1cm

\address
{$^a$  Department of Physics, Sharif University of Technology, P.O.
Box 11365-9161, Tehran, Iran.}

\maketitle

%%%%%%%%%%%%%%%%%%%%%%%%%%%%%%%%%%%%%%%%%%%%%%%%%%%%%%
%ABSTRACT
%%%%%%%%%%%%%%%%%%%%%%%%%%%%%%%%%%%%%%%%%%%%%%%%%%%%%%
\begin{abstract}
The Heisenberg spin chain (with nearest neighbor interaction) in an
external magnetic field, is defined by 3 coupling constants (after
we re-scale the energy by a multiplicative constant). We show that
on a particular two dimensional hypersurface, the ground state and
all the correlation functions can be determined exactly and in
compact form. This ground state has a very interesting property: all
the pairs of spins are equally entangled with each other. In this
last respect the results may be of interest for engineering
long-range entanglement in experimentally realizable finite arrays
of qubits.

PACS: 03.67.-a, 03.65.Ud, 03.67.Mn, 05.50.+q
\end{abstract}
\hspace{.3in}
\newpage]
A basic problem in condensed matter physics is to find the ground
state of a given Hamiltonian embodying the interactions of a many
body system. A prototype of such a system is the Heisenberg spin 1/2
chain, described by the following Hamiltonian
\begin{eqnarray}\label{H}
   H &=&\sum_{i=1}^N J_x \sigma_{x,i}\sigma_{x,{i+1}}+J_y \sigma_{y,i}\sigma_{y,i+1}\cr
  &+&J_z \sigma_{z,i}\sigma_{z,i+1}+B \sigma_{x,i},
\end{eqnarray}
where, $\sigma_{a}$'s are the Pauli operators, $J_a$'s are the
coupling strengths and $B$ is the external magnetic field. There is
a long history of attempts for finding exact solutions of this model
at some special points or lines in this parameter space
\cite{korepin}. The matrix product formalism, originally introduced
and developed in \cite{aklt,mpsbasic}, has been recently revived
\cite{mpscirac,ver1} mainly due to the work in quantum information
community, where the emphasis is on the properties of many body
states, like their entanglement or quantum correlations
\cite{fan,osterloh,occoner}. In this formalism, one first constructs
a many body state and then finds the Hamiltonian for which this
state is an exact ground state. As is always the case, when we
reverse a difficult problem (in this case finding the ground state
of an interacting spin system), the difficulty shows up in some
other form in some other place: except for very rare cases,
\cite{aklt} the Hamiltonians which are found are not usually simple
and of wide interest to condensed matter physicists
\cite{stochastic}. In this letter, we show for the first time that
the Heisenberg spin 1/2 chain can be solved exactly and in compact
form on a two dimensional surfaces defined by
\begin{eqnarray}\label{BJ}
    J_x &=&-J+\frac{1+g^2}{2} ,\ \ J_y=-\eta J+g, \cr \  J_z&=&-\eta
    J-g \ \ \ \ , \ \ \ \  B=\epsilon(g^2-1),
\end{eqnarray}
in which $(\epsilon,\eta)=\pm (1,\pm 1)$ are two discrete
parameters, and $g$ and ($J>0$) are two continuous parameters.
Unlike the ${\rm xxx}$ or ${\rm xxz}$ Heisenberg anti-ferromagnetic
chains whose solutions are implicitly given via the solution of
Bethe ansatz equations, the ground states of these models can be
determined quite explicitly and expressed in terms of simple
functions. Yet as we will see these ground states are quite rich in
their properties. We will calculate the spin correlation functions
exactly and show that singularities in the thermodynamic limit
develop at $g=0$, a property which has been called MPS-Quantum Phase
Transition in \cite{mpscirac}, to distinguish them from  known
examples of QPT's \cite{sachdev}. \\

\begin{figure}
\epsfxsize=7.5truecm\epsfbox{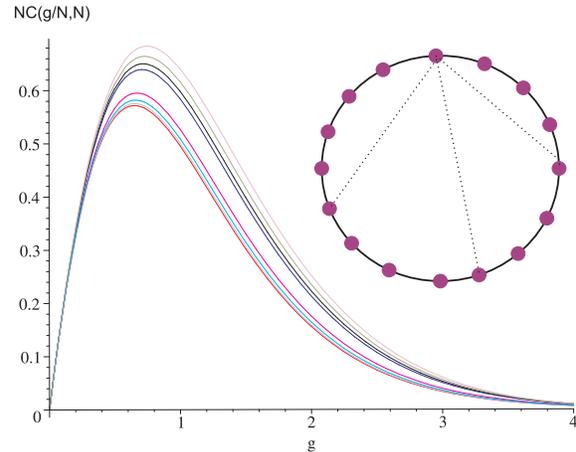} \narrowtext \caption{In the
MPS ground state of (\ref{H}), all the pairs of spins are equally
entangled with each other. The curves show scaled concurrence for
rings of size(from top to bottom) 6,7,8,9,10,20,30,40,and
50.}\label{ring}
\end{figure}

We will also show that, these ground state have the very interesting
property that all the pairs of spins have equal entanglement with
each other. This is a very desirable situation for quantum
information processing, both theoretically and experimentally, i.e.
an array of qubits in which there are long range entanglement,
figure (\ref{ring}).
%%%%%%%%%%%%%%%%%%%%%%%%%%%%%%%%%%%%%%%%%%%%%%%%%%%%%%%%%%%%%%%%%%%%%%%%%
%%%%%%%%%%%%%%%%%%%%%%%%%%%%%%%%%%%%%%%%%%%%%%%%%%%%%%%%%%%%%%%%%%%%%%%
Let us briefly review the MPS formalism. On a ring of $N$ sites of
$d-$level particles, a state is called a matrix product state if
there exist matrices $A_i, i=0,\cdots, d-1$ (of dimension $D$) such
that
\begin{equation}\label{mat}
    \psi_{i_1,i_2,\cdots
    ,i_N}=\frac{1}{\sqrt{Z}}tr(A_{i_1}A_{i_2}\cdots A_{i_N}),
\end{equation}
where $Z$ is a normalization constant given by $
    Z=tr(E^N)
$ in which $  E:=\sum_{i=0}^{d-1} (A_i^*\otimes A_i). $ The state
(\ref{mat}) is reflection symmetric if there exists a matrix $\Pi$
such that $A_i^T=\Pi A_i\Pi^{-1}$ (where $T$ means transpose) and
time-reversal invariant if there exists a matrix $V$ such that
$A_i^*=VA_iV^{-1}$.  All the correlation functions can be calculated
exactly. For example, for a local observable $O$, one finds
\begin{equation}\label{1point}
    \la \Psi|O(k)|\Psi\ra = \frac{tr(E^{k-1}E_O E^{N-k})}{tr(E^N)},
\end{equation}
where $E_O := \sum_{i,j=0}^{d-1}\la i|O|j\ra A_i^*\otimes A_j.$ In
the thermodynamic limit ($N\rightarrow\infty$), only the
eigenvector(s) corresponding to the eigenvalue $\lambda_{max}$ of
$E$ with the largest absolute value matters and any level-crossing
in this  eigenvalue leads to a discontinuity in
correlation functions.\\
Given a matrix product state, the reduced density matrix of $k$
adjacent sites is given by
$$
    \rho_{i_1\cdots i_k,j_1\cdots j_k}=\frac{tr((A_{i_1}^*\cdots A_{i_k}^*\otimes A_{j_1}\cdots A_{j_k})E^{N-k})}{tr(E^N)}.
$$
This density matrix has at least $d^k-D^2$ zero eigenvalues. To see
this, suppose that we can find complex numbers $c_{i_1\cdots i_k}$
such that
\begin{equation}\label{ccA}
    \sum_{j_1,\cdots, j_k=0}^{d-1}c_{j_1\cdots j_k}
A_{j_1}\cdots A_{j_k}=0. \end{equation} This is a system of $D^2$
equations for $d^k$ unknowns which has at least $d^k-D^2$
independent solutions. Any such solution gives a null eigenvector of
$\rho$. Thus for the density matrix of $k$ adjacent sites to have a
null space, it is sufficient (but not necessary) that $ d^k\
>\ D^2. $ Let the null space of the reduced density matrix be
spanned by the orthogonal vectors $|e_{\a}\ra, \a=1,\cdots, s$, then
we can construct the local Hamiltonian acting on $k$ consecutive
sites as
$$
    h:=\sum_{\a=1}^s J_{\a} |e_{\a}\ra\la e_{\a}|,
$$
where $J_{\a}$ are positive constants. The total Hamiltonian on the
chain will then be given by the positive operator $ \ \
H=\sum_{l=1}^N h_{l, l+k},\ $ where $h_{l,l+k}$ is the embedding of
$h$ into sites $l$ to $l+k$ of the chain. The state $|\psi\ra$ will
then be a ground state of $H$.
%%%%%%%%%%%%%%%%%%%%%%%%%%%%%%%%%%%\section{Spin chains with nearest neighbor interactions}
Equation ($d^k>D^2$) puts a stringent requirement on the dimensions
of the matrices used in construction of a matrix product state. When
dealing with spin $1/2$ with nearest-neighbor interactions, for
which $d=2$ and $k=1$, it appears that the only admissible dimension
for the matrices $A_0$ and $A_1$ is $D=1$, leading to a product
state. However, it is crucial to note that the condition $d^2>D^2$
is only a sufficient and not a necessary
condition for the density matrix $\rho$ to have a null space. \\
To proceed with our construction, we require that the state satisfy
some natural symmetries, i.e. spin-flip symmetry which, in the
language of matrix product formalism, means that there is a matrix
$X$ such that
$$
XA_{0}X^{-1}=\epsilon A_{1},\ \ \ XA_{1}X^{-1}=\epsilon A_{0},
$$
where $\epsilon^2=1$. Working in the basis where $X=\sigma_z$, we
find the general form of the matrices $A_0$ and $A_1$:
$$
    A_0=\left(\begin{array}{cc} a & b \\ c & d \end{array}\right)\h  A_1=\epsilon\left(\begin{array}{cc} a & -b \\ -c & d
    \end{array}\right).
$$
Although these two matrices are not symmetric, the state constructed
from them is symmetric under parity, since there is a matrix
$\Pi=\left(\begin{array}{cc} b & 0
\\ 0 & c
\end{array}\right)$
with the property
$$
    \Pi A_0^t\Pi^{-1}=A_0, \h   \Pi A_1^t\Pi^{-1}=A_1.
$$
We now consider the matrix equation (\ref{ccA}) which in the present
case is
\begin{equation}\label{cA}
c_{00}A_0^2 + c_{01}A_0A_1+c_{10}A_1A_0 + c_{11}A_1^2=0.
\end{equation}
This is a set of linear equations for the four coefficients
$c_{ij}$, which can be written as a matrix equation $MC=0$, leading
to a non-zero solution when
$$
    det(M)\equiv 16 b^2 c^2 (a-d)^2 (a+d)^2=0.
$$
Thus we will find non-trivial models, for $a=d$ or $a=-d$. The
models with $b=0$ or $c=0$ are not symmetric under parity, since in
these cases the matrix $\Pi$ will not be invertible. We can always
re-scale the matrices by a constant factor without affecting the
matrix product state, so we set $a=1$ and use a subsequent gauge
transformation $A_i\ro SA_iS^{-1}$ with $S=\left(\begin{array}{cc} c
& 0
\\ 0 & 1 \end{array}\right)$, to set $c=1$. Therefore we are left with the following four classes of models
defined by the matrices
\begin{equation}\label{modelA}
    A_0=\left(\begin{array}{cc} 1 & g \\ 1 & \eta\end{array}\right)\h A_1=\epsilon\left(\begin{array}{cc} 1 & -g \\ -1 &
    \eta\end{array}\right),
\end{equation}
where $g$ is a continuous parameter and $(\epsilon,\eta)=\pm (1,\pm
1) $. The four types of models are distinguished by the values of
the pair $(\epsilon, \eta)$. The eigenvalues of the matrix
$E=A_0\otimes A_0+A_1\otimes A_1$ are $ 2(\eta \pm g),\ \ 2(1\pm g)\
.$ The correlation functions can be derived from (\ref{1point}).
Consider for definiteness, the case $\eta=1$. The magnetization per
site is found from (\ref{1point}) to be
$$
    \la \sigma_y\ra = \la \sigma_z\ra = 0,\h \la \sigma_x\ra =
    \epsilon {u}\frac{1+u^{N-2}}{1+u^N},
$$
where $u:=\frac{1-g}{1+g}$
 and the correlation functions $G_a(1,r):= \la
\sigma_{a,1}\sigma_{a,r}\ra $ are similarly found to be as follows:
\begin{eqnarray}\label{correlations+1}
&&G_x(1,r) =  \frac{u^2+u^{N-2}}{1+u^N}, \cr  && G_y(1,r) =
\frac{u^{N-2}(u^2-1)}{1+u^{N}}, \ \  G_z(1,r) =
\frac{1-u^2}{1+u^{N}}.
\end{eqnarray}
These correlation functions satisfy the following relations:
\begin{equation}\label{condition} G_x+G_y+G_z=1\ \ ,\ \ (1-G_z)(1-G_y)=\la \sigma_x\ra^2.
\end{equation} In the thermodynamic limit ($N\ro \infty$), discontinuities
develop in these correlation functions at $g=0$. For example, the
magnetization per site will be
$$
    \la \sigma_x\ra = \epsilon \frac{1+|g|}{1-|g|},
$$
%%%%%%%%%%%%%%%%%%%%%%%%%%%%%%%%%%%%%%%%%%%%%%%%%%%%%%%%%%%%%%%%%%%%%%%%%
\begin{figure}\label{newXg}
\centering
    \includegraphics[width=5cm,height=3cm,angle=0]{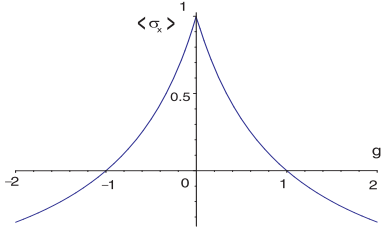}\vspace{1cm}
    \caption{(Color Online)The
magnetization in the $x$ direction as a function of $g$,
($\epsilon=1$).}
\end{figure}
%%%%%%%%%%%%%%%%%%%%%%%%%%%%%%%%%%%%%%%%%%%%%%%%%%%%%%%%%%%%%%%%%%%%%%%
and displayed in figure (2). Interestingly, the spins align
themselves opposite to the magnetic field. To understand this, let
us set for definiteness $\epsilon=\eta=1$ and consider the two-body
Hamiltonian $H_2$, after subtracting the ferromagnetic term
$-J{\sigma} \cdot {\sigma}$ for which any product state
$|\phi\ra|\phi\ra$ is an eigenstate. For $|g|\gg 1$ the remaining
Hamiltonian tends to $ H_2\sim
g^2(\frac{1}{2}\sigma_{1x}\sigma_{2x}+\sigma_{1x}+\sigma_{2x})$ ,
with the ground state $|x-\ra|x-\ra$ and for $|g|\ll 1$ it tends to
$H_2\sim
(\frac{1}{2}\sigma_{1x}\sigma_{2x}-\sigma_{1x}-\sigma_{2x})$, with
the ground state $|x+\ra|x+\ra$. It is thus the combination of the
anti-ferromagnetic interaction in the $x$ direction and the magnetic
field which align the spins opposite to the magnetic field. Even
when we tune the coupling $J$ so that there is no $xx$ coupling,
(i.e. $J=(1+g^2)/2$ ) this anti-alignment happens.  To understand
this, consider $g\approx 1$, where $H_2\approx -
\sigma_{1z}\sigma_{2z}+B(\sigma_{1x}+\sigma_{2x})$ with $B$ very
small. The ground state of this Hamiltonian is $|\psi\ra\approx
|\phi_+\ra-B|\psi_+\ra$, where $|\psi_+\ra$ and $|\phi_+\ra$ are
Bell states. One then finds that $\la
\sigma_{1x}+\sigma_{2x}\ra=-4B$, confirming figure (2). \\

In order to see how the Hamiltonian is constructed, we solve
equations (\ref{cA}) which in view of (\ref{modelA}) take the form
\begin{eqnarray}
  (1+g)(C_{00}+C_{11})+\epsilon(1-g)(C_{01}+C_{10}) &=& 0 \\
  (1+\eta)(C_{00}-C_{11})-\epsilon(1-\eta)(C_{01}-C_{10}) &=& 0.
\end{eqnarray}
It is easy to verify that the solution space is determined by the
following two un-normalized vectors,
\begin{eqnarray*}
    |e_1\ra &=& (1+\eta)|\psi_-\ra +(1-\eta)|\phi_-\ra\cr
    &&\cr
    |e_2\ra &=& (1+g)|\psi_+\ra-\epsilon(1-g)|\phi_+\ra,
\end{eqnarray*}
where $|\psi_{\pm}\ra=\frac{1}{\sqrt{2}}(|0,1\ra\pm|1,0\ra)$ and
$|\phi_{\pm}\ra=\frac{1}{\sqrt{2}}(|0,0\ra\pm|1,1\ra)$ are Bell
states. Under spin flip the above states transform as
$|e_{1,2}\ra\lo \mp |e_{1,2}\ra$. The final local Hamiltonian will
be given by $ h=J|e_1\ra\la e_1|+|e_2\ra\la e_2|, $ where $J$ is a
non-negative parameter and we have used the freedom for rescaling
the couplings of the Hamiltonian to set one of the parameters equal
to 1. In view of the symmetry property of the vectors under spin
flip, this Hamiltonian will be symmetric under spin flip. Expressing
the above operator in terms of Pauli operators and subtracting
constant terms, we find the total Hamiltonian which is written in
(\ref{H}) and (\ref{BJ}).  \\
Note that the models with $\epsilon = \pm 1$ (sign of the magnetic
field) are related by local $\pi $-rotations of spins around the $z$
axis where ($\sigma_{x,y}\ro -\sigma_{x,y}$).  Also, the models with
$\eta=\pm$  are related to each other by simultaneous rotations
$R_x(\frac{\pi}{2})\otimes R_x(\frac{-\pi}{2})$ of spins on adjacent
sites, under which we have
($\sigma_{z,i}\sigma_{z,i+1}\rightleftharpoons
-\sigma_{y,i}\sigma_{y,i+1})$. This is, of course, possible only
when $N$ is even \cite{Nodd}. \\

The explicit form of such a ground state can also be determined. For
$\eta=1$, the matrices $A_0$ and $A_1$ commute. By a similarity
transformation which does not change the state (\ref{mat}), both the
matrices are made diagonal,
$$
    A_0=\left(\begin{array}{cc} 1+\sqrt{g}& 0 \\ 0 &
    1-\sqrt{g}\end{array}\right)\ ,   A_1=\left(\begin{array}{cc} 1-\sqrt{g}& 0 \\ 0 & 1+\sqrt{g}\end{array}\right)
$$
and the MPS state (\ref{mat}) will be given by
\begin{equation}\label{groundstate}
|\Psi\ra_{\eta=1} = \frac{1}{\sqrt{Z}}(|\phi_+\ra^{\otimes
N}+|\phi_-\ra^{\otimes N}),
\end{equation}
where $$ |\phi_{\pm}\ra =(1\pm \sqrt{g})|0\ra + (1\mp
\sqrt{g})|1\ra,$$ and $Z=2^{N+1}((1+g)^N+(1-g)^N)$. These
expressions are valid for all values of $g$, provided that we
replace $\sqrt{g}\lo i\sqrt{-g}$ when we consider negative values of
$g$. Note that  $\la \phi_+|\phi_-\ra=2(1-g)$.  One can indeed check
that the separate product states are ground states of $H$, (i.e. the
local Hamiltonian acting on two adjacent sites, when added by a
suitable constant, annihilates $|\phi_{\pm}\ra^{\otimes 2}$).
However the advantage of the MPS state we have constructed
$|\phi_{+}\ra^{\otimes N}+ |\phi_-\ra^{\otimes N}$ or the other
state $|\phi_{+}\ra^{\otimes N}- |\phi_-\ra^{\otimes N}$ is that
they are invariant under spin-flip transformation $\sigma_x^{\otimes
N}$. Thus even if the couplings of the Hamiltonian are not tuned as
exactly as in (\ref{BJ}), and are perturbed a little bit, first
order perturbation theory guarantees that one of these entangled
states and not the product states, will be the
unique grounds state of $H$.\\

We now come to the entanglement properties of the state
(\ref{groundstate}). At $g=1$ when $\la \phi_+|\phi_-\ra=0$, the
state becomes a standard GHZ state, $\frac{1}{\sqrt{2}}(|0\cdots
0\ra+|1\cdots 1\ra)$. For other values of $g$, when $|\phi_+\ra$ and
$|\phi_-\ra$ are no longer orthogonal, it can be named a generalized
$GHZ$ state. Obviously such a state induces equal entanglement
between any two spins regardless of their distance.  To calculate
this entanglement we determine the reduced two particle density
matrix and use Wootters formula \cite{wootters}, with the result
\cite{conc}:
$$
   C=\frac{4|g|}{|(1+g)^N+(1-g)^N|}|1-|g||^{N-2}.
$$
Thus although the ring is not totally connected, the mutual
entanglement of all pairs are equal and independent of their
distances. Looking at the $N\gg 1$ limit, one can obtain the
relation
\begin{equation}\label{scaling}
    NC(\frac{g}{N},N)\approx \frac{2|g|e^{-|g|}}{\cosh g}.
\end{equation}
One can interpret the left hand side as the total mutual
entanglement of a spin with all the other spins, and the above
equation as a universal scaling relation for this total
entanglement. \\

For the case $\eta=-1$, we use the transformation
$U:=R_x(\frac{\pi}{2})\otimes R_x(\frac{-\pi}{2})$ on adjacent sites
which transforms $H(\eta=1)$ to $H(\eta=-1)$ and act on
(\ref{groundstate}) by $U^{\otimes \frac{N}{2}}$ to obtain the
ground state as
\begin{equation}\label{groundstate2}
|\Psi\ra_{\eta=-1} =
\frac{1}{\sqrt{Z}}(|\chi_+\ra|\chi_-\ra)^{\otimes
\frac{N}{2}}+(|\chi_-\ra|\chi_+\ra)^{\otimes \frac{N}{2}},
\end{equation}
where $$
    |\chi_{\pm}\ra
    =(1+\sqrt{g})|y,\pm\ra\ \pm\ i(1-\sqrt{g})|y,\mp\ra,
$$
in which  $|y,\pm\ra$ denote the eigenstates of $\sigma_y$. An
alternative way for deriving this ground state and indeed the reason
for its simple structure, is to note that for $\eta=-1$, although
the matrices $A_0$ and $A_1$ do not commute, the pairs of matrices
corresponding to Bell states, defined by
$$\Phi_{mn}:=\frac{1}{\sqrt{2}}(A_{0}A_m+(-1)^nA_{1}A_{1+m}), \ \ m,n=0,1$$
commute with each other. The reader can verify that the states
$|\chi\ra_{\pm}|\chi_{\mp}\ra$ are indeed linear combinations of the
Bell states $\phi_{00}\equiv\phi_+$ , $\phi_{01}\equiv\psi_-$ and $\phi_{10}\equiv\psi_+$,
making the state (\ref{groundstate2}) a linear superposition of strings of various Bell states on adjacent sites.\\
In summary we have introduced a two-parameter family of spin 1/2
${\rm xyz}$ Heisenberg chains, with nearest neighbor interactions in
an external magnetic field, for which ground states and all
correlation functions can be calculated exactly.  These states have
two very interesting properties: first, they undergo a discontinuous
(quantum phase) transition as one of the parameters passes a
critical point, which stimulates further exploration of MPS-quantum
phase transition in a set of important exactly solvable models.
Second, they have the property that all the pairs of spins are
equally entangled with each other. This makes them good candidates
for engineering long-range entanglement in experimentally realizable
arrays of qubits or spin systems. This study can be extended in
several directions, including generalization to open chains, finding
the excitations, perturbing them to explore more models, and actual
engineering of such chains for small number of qubits for
information processing tasks. We thank David Gross, of Imperial
college, London for a very stimulating email correspondence which
led to substantial improvement of this paper and also A. Langari and
M. R. Rahimitabar, for their valuable comments.  Corresponding
author, V. Karimipour, vahid@sharif.edu.
{}

\begin{thebibliography}{99}
\bibitem{korepin} V. E. Korepin and O. I. Patu, preprint,
Cond-Mat/0701491.
\bibitem{aklt} I. Affleck, T. Kennedy, E.H. Lieb and H. Tasaki,
Phys. Rev. Lett. {\bf 59}, 799 (1987).
\bibitem{mpsbasic} M. Fannes, B. Nachtergaele and R. F. Werner, Europhys. Lett. {\bf 10} 633, (1989), A. Klumper, A. Schadschneider, J. Zittartz,
J. Phys. A {\bf 24} L955 (1991); Z. Phys. B, {\bf 87}, 281 (1992).
\bibitem{mpscirac} M. M. Wolf, G. Ortiz, F. Verstraete, J. I. Cirac,
Phys. Rev. Lett.{\bf 97}, 110403 (2006); D. Peres Garcia, et al,
quant-ph/0608197.
\bibitem{ver1}F. Verstraete, M.A. Martin-Delgado, J.I. Cirac, Phys. Rev. Lett. 92, 087201
(2004);  F. Verstraete, M. Popp, J.I. Cirac, Phys. Rev. Lett. 92,
027901 (2004);  F. Verstraete, J.I. Cirac, J.I. Latorre, E. Rico,
M.M. Wolf, Phys.Rev.Lett. 94 (2005) 140601.
\bibitem{fan} H. Fan, V. E. Korepin and V. Roychowdhury, Phys. Rev.
Letts., {\bf 93}, 227203 (2004); A. R. Its, B.-Q. Jin, and V. E.
Korepin, Jour. Phys. A. Math. Gen. {\bf 38}, 2975 (2005).
\bibitem{osterloh} A. Osterloh, L. Amico, G. Falci and R. Fazio, Nature 416, 608
(2002); T. J. Osborne, and M. A. Nielsen, Phys. Rev. A 66, 032110
(2002); M. C. Arnesen, S. Bose, and V. Vedral, Phys. Rev. Lett. {\bf
87}, 277901 (2001); M. Cozzini, Radu Ionicioiu, and Paulo Zanardi,
Quantum fidelity and quantum phase transition in matrix product
states, Cond-mat/0611727.
\bibitem{occoner} K. M. O'Connor and W. K. Wootters, Phys. Rev. A 63 (2001)
052302; M. Asoudeh, and V. Karimipour, Phys. Rev. A {\bf 70}, 052307
(2004).
\bibitem{stochastic} Here we are exlcuding the application of the MPS formalizm in stochastic systems,
where ground state of non-hermitian matrices represent steady state
of a stochastic process, B. Derrida, M.R. Evans, V. Hakim, V.
Pasquier, J. Phys. A, {\bf 26}, 1493 (1993), V. Karimipour, Phys.
Rev. E59, 205 (1999).
\bibitem{sachdev} S. Sachdev, {\it Quantum Phase Transitions}
(Cambridge University Press, Cambridge, 1999).
\bibitem{wootters} W. K. Wootters, Phys. Rev. Letts. {\bf 80}, 2245 (1998).
\bibitem{conc} With conditions (\ref{condition}), the matrix $\rho\tilde{\rho}$ has only two
non-zero eigenvalues given by $
    \lambda_{1,2}=\frac{1}{4}(G_y\pm G_z)^2,
$ which gives $
    C=|\sqrt{\lambda_1}-\sqrt{\lambda_2}|.
$
\bibitem{Nodd} For $\eta=-1$ and $N=odd$ such an MPS does not exist, i.e.
$\Psi$ is identical to zero. One can prove this by induction using
the facts that $tr(A_0)=tr(A_1)=0$, and $A_0^2, A_1^2$ and
$A_0A_1+A_1A_0$ are all proportional to identity.
\end{thebibliography}
\end{document}